# HYPIC: A fast hybrid EM PIC-MCC code for ion cyclotron resonance energization in cylindrical coordinate system

Mingyang Wu (吴明阳)[1], Andong Xu (许安冬)[1], Chijie Xiao (肖池阶)[1,*]

[1]State Key Laboratory of Nuclear Physics and Technology, School of Physics, Fusion Simulation Center, Peking University, Beijing 100871, China

[*]Corresponding author: cjxiao@pku.edu.cn

## Abstract

Ion cyclotron resonance energization (ICRE) such as ion cyclotron resonance heating (ICRH) is widely applied to magnetic confinement fusion and high-power electric propulsion. Since ICRE involves cyclotron resonance processes, a kinetic model is required. Both conventional particle-in-cell (PIC) simulations and solving the Boltzmann equation require enormous computation and memory. The hybrid simulation incorporating of adiabatic electrons and PIC ions allows both a substantial reduction in computation and the inclusion of cyclotron resonance effects. Under the adiabatic electron approximation, we have developed a two-dimensional $(r, z)$ hybrid electromagnetic (EM) PIC-MCC (Monte-Carlo collision) simulation program, named HYPIC. The advantages of HYPIC are the inclusion of ion kinetic effects, electrostatic (ES) and EM effects, and collisional effects of ions and electrons, with a small computation. The HYPIC program is able to fast simulate the antenna-plasma interactions and the ion cyclotron resonance energization and/or ion cyclotron resonance heating processes in linear devices, such as high-power electric propulsion, magnetic mirror, and field-reversed-configuration (FRC), etc.

**Program summary**

*Program title:* HYPIC

*Licensing provisions:* BSD 3-clause

*Programming language:* Fortran

*Nature of problem:* The code solves the electromagnetic field distribution in wave-plasma interactions and the motion of ions in the presence of electrostatic, electromagnetic, background magnetic fields, and collisions in two-dimensional (2D) $(r,z)$ cylindrical coordinate system. The code is able to fast simulate the antenna-plasma interactions, ion cyclotron resonance energization and/or ion cyclotron resonance heating processes in linear devices, such as high-power electric propulsion, magnetic mirror, and field-reversed-configuration, etc.

*Solution method:* The PIC method with adiabatic electrons is used to fast simulate of the motion of ions by 4-stage Runge–Kutta time integral. The frequency-domain finite-difference method to solve Maxwells equations allows for a fast solution of radio-frequency wave-plasma interactions. Monte Carlo collisions are used to deal with collisions of electrons and ions.

*Additional comments including restrictions and unusual features:* The current version only supports solving axisymmetric 2D problems.

# 1 Introduction

Ion cyclotron resonance energization (ICRE) such as ion cyclotron resonance heating (ICRH) is widely applied to both magnetic confinement fusion [1-5] and high-power electric propulsion [6, 7]. The advantages of ICRE over electron cyclotron resonance heating (ECRH) and neutral beam injection (NBI) are the low cost of the radio-frequency (RF) source and the ability to directly energize ions [8]. The challenge for ICRE is to increase the efficiency of energy conversion between RF waves and plasmas [9]. Since ICRE involves cyclotron resonance processes, it is necessary to use kinetic methods such as conventional particle-in-cell (PIC) simulations and solving the Boltzmann equation, requiring huge computation and memory [10-15]. The hybrid simulation incorporating of adiabatic electrons and PIC ions allows both a substantial reduction in computation and the inclusion of cyclotron resonance effects. Lin et al used a hybrid PIC-MCC (Monte-Carlo Collision) model with adiabatic electron approximation to study ICRE in high-power electric propulsion[11, 16, 17]. However, Lin et al. did not consider electron collisions and artificially set the electron temperature as a constant. Ion-electron collisions increase the electron temperature proportional to the electrostatic (ES) field which affects the ICRE processes, plasma confinement, and propulsion efficiency. Thus, we add the MCC of electrons to the models of Lin et al [11, 16, 17] and develop the HYPIC program able to study the collisional effects of electron on ICRE. We have carefully verified the HYPIC program and given a detailed example of ICRE in magnetic mirror [2, 5]. The advantages of the HYPIC program are the inclusion of (1) the ion kinetic effects, (2) ES and electromagnetic (EM) effects, (3) the collisional effects of ions and electrons, and (4) with a small computation, e.g., the example given in section 3 consuming 20 hours by one thread. The HYPIC program can be used to fast simulate the antenna-plasma interactions and the ICRE processes in linear devices, such as high-power electric propulsion, magnetic mirror, and field-reversed-configuration (FRC), etc.

This paper is organized as follows. Section 2 presents the simulation model including the governing equations, numerical methods and benchmarks. The applications of HYPIC and a detail example are presented in Section3. In Section 4, we summarize and discuss.

## 2 Simulation model
### 2.1 Governing equations

We use a hybrid EM PIC-MCC model that the momentum equation of ions is solved directly with the adiabatic electron approximation, and the collisions of electrons and ions are solved through the MCC method [18]. The total electric field $E$ is decomposed into the ES field $E_s$ and the induced electric field $E_{mf}$, i.e., $E = E_s + E_{mf}$. The ES field $E_s$ can be obtained from the Boltzmann distribution. The induced electric field $E_{mf}$, on the other hand, is obtained by solving Maxwells equations in frequency domain. Details are provided below.

The momentum equation of ions solved in HYPIC is

$$m\frac{dv}{dt} = qE + qv \times B_0 + F_c, \quad (1)$$

where the magnetic field contains only the background magnetic field $B_0$ since the induced magnetic field is much smaller than the background magnetic field. Solving Amperes law

$$\nabla \times B_0 = \mu J_c, \quad (2)$$

yields the background magnetic field, where $\mu$ is the permeability and $J_c$ is the current density of coils [19]. The change of momentum $F_c$ due to collisions and is solved by the MCC method. Four types of collisions are considered, including ion-ion (i-i), ion-electron (i-e), electron-electron (e-e), and electron-ion (e-i) collisions. The adiabatic electron approximation assuming the mass of electron $m_e = 0$ is applied, then the momentum equation of electrons becomes

$$\boldsymbol{E}_s = T_e \frac{\nabla n_e}{n_e} + \boldsymbol{u}_e \times \boldsymbol{B}_0, \quad (3)$$

where $n_e, T_e, \boldsymbol{u}_e$ are the electron density, electron temperature, and fluid velocity of electrons, respectively. In the direction parallel to the magnetic field, the Hall term (the second term on the right-hand side of the equation (3)) is equal to 0. In the direction perpendicular to the magnetic field, both strong and weak, order of magnitude estimation yields that the pressure gradient term (the first term on the right-hand side of the equation (2)) is much larger than the Hall term. For example, by choosing the electron temperature $T_e = 100$ eV, radial plasma radius $r_p = 0.1$ m, perpendicular diffusion coefficient $D_\perp \approx 3$ m$^2$/s and magnetic field $B_0 = 0.4$ T, we have $u_{e\perp} = D_\perp/r_p = 30$ m/s, and $\nabla_\perp p_e/en_e = T_e/r_p = 10^3$ V/m $\gg u_{e\perp}B_0 = 12$ V/m. Therefore, the Hall term can be neglected either perpendicular or parallel to the direction of the magnetic field and then the equation (2) can be simplified to

$$\boldsymbol{E}_s = T_e \frac{\nabla n_e}{n_e}. \quad (4)$$

Macroparticle, an aggregate of $N_m$ real particles (charge and mass), is usually used in PIC simulations. Because the macroparticle number $N_m$ is canceled out in equation (4) under the adiabatic electron approximation, we can use the real ion mass and charge in the equation (2). Therefore, the adiabatic electron approximation not only eliminates the numerical instability caused by electrons, but also avoids the numerical error caused by large $N_m$ when calculating the electrostatic field. However, we still need to determine the macroparticle number $N_m$ to obtain plasma density and power consumption. The $N_m$ is calculated by $N_m = \frac{1}{N_p}\iiint n_{i0}(r,z)rdrd\theta dz$, where $n_{i0}(r,z)$ is the initial plasma density distribution and $N_p$ is the number of ions at the initial time. The ES field is proportional to the electron temperature according to equation (4), and here we determine the electron temperature $T_e$ by MCC (collisions associated with electrons). At the initial moment, we give equal numbers of electrons and ions, after which the velocities of electron change due to collisions (the electron positions not evolving). The electron temperature $T_e$ with unit of eV at any instant can be calculated from the mean kinetic energies of all electrons, i.e., $T_e = 2E_{ke}/3$. We assume that the electron temperature transports fast enough so that the uniform electron temperature can be adopted.

The Maxwells equations in the frequency domain in the plasma region is

$$\nabla(\nabla \cdot \boldsymbol{E}_{mf}) - \nabla^2 \boldsymbol{E}_{mf} - \frac{\omega^2}{c^2}\overleftrightarrow{\varepsilon} \cdot \boldsymbol{E}_{mf} = i\omega\mu_0 \boldsymbol{J}_a. \tag{5}$$

In the vacuum region, with simplification of $\nabla \cdot \boldsymbol{E}_{mf} = 0$, we have

$$-\nabla^2 \boldsymbol{E}_{mf} - \frac{\omega^2}{c^2}\varepsilon_0 \boldsymbol{E}_{mf} = i\omega\mu_0 \boldsymbol{J}_a, \tag{6}$$

where $\boldsymbol{J}_a$ is the antenna current density and $\overleftrightarrow{\varepsilon} = I + i\overleftrightarrow{\sigma}/(\varepsilon_0\omega)$ is the plasma dielectric tensor. Since the ion current is much smaller than the electron current, only the contribution of the electron current is considered in the plasma conductivity $\overleftrightarrow{\sigma}$,

$$\overleftrightarrow{\sigma} = \frac{\varepsilon_0 \omega_{pe}^2}{(\nu_e - i\omega)((\nu_e - i\omega)^2 + \omega_{ce}^2)}\begin{bmatrix} (\nu_e - i\omega)^2 + \omega_{cr}^2 & -(\nu_e - i\omega)\omega_{cz} & \omega_{cr}\omega_{cz} \\ (\nu_e - i\omega)\omega_{cz} & (\nu_e - i\omega)^2 & -(\nu_e - i\omega)\omega_{cr} \\ \omega_{cr}\omega_{cz} & (\nu_e - i\omega)\omega_{cr} & (\nu_e - i\omega)^2 + \omega_{cz}^2 \end{bmatrix}, \tag{7}$$

with the Langmuir oscillation frequency $\omega_{pe} = \sqrt{e^2 n_e/(\varepsilon_0 m_e)}$, the electron cyclotron frequency $\omega_{cr} = eB_r/m_e, \omega_{cz} = eB_z/m_e, \omega_{ce} = eB_0/m_e$, and the electron collision frequency $\nu_e$ [20]. In order to save computational resource, we use the finite-difference frequency-domain (FDFD) instead of the finite-difference time-domain (FDTD) methods used in the PHD v1.1 program [21]. By the Fourier transform in time and azimuth, we have $f(r,\theta,z,t) = \sum_{m=-\infty}^{+\infty} f(r,z)e^{im\theta - i\omega t}\boldsymbol{e}_\theta$ ($f = \boldsymbol{E}_{mf}, \boldsymbol{J}_a$).

The governing equations of the HYPIC program are summarized as

$$\begin{cases} m\dfrac{d\boldsymbol{v}}{dt} = q\boldsymbol{E} + q\boldsymbol{v} \times \boldsymbol{B_0} + \boldsymbol{F}_c \\ n_e = n_i \\ \boldsymbol{E}_s = T_e \dfrac{\nabla n_e}{n_e} \\ \nabla(\nabla \cdot \boldsymbol{E}_{mf}) - \nabla^2 \boldsymbol{E}_{mf} - \dfrac{\omega^2}{c^2}\overleftrightarrow{\varepsilon} \cdot \boldsymbol{E}_{mf} = i\omega\mu_0 \boldsymbol{J}_a \text{ in plamsa} \\ -\nabla^2 \boldsymbol{E}_{mf} - \dfrac{\omega^2}{c^2}\overleftrightarrow{\varepsilon} \cdot \boldsymbol{E}_{mf} = i\omega\mu_0 \boldsymbol{J}_a \text{ in vacuum} \end{cases}. \tag{8}$$

## 2.2 Numerical method

The momentum equation (1) are solved in a three-dimensional (3D) Cartesian coordinate system by 4-stage Runge–Kutta (RK4) method [22]. In PIC simulations, since particles are defined in continuum space in both position and velocity and fields are defined on grids, the interaction of particles with fields occurs through linear interpolation scheme [10]. Here we consider azimuthal mode number m = 0, thus a 2D $(r,z)$ interpolation, usually called area weighting, can be performed in

$$f_p = \frac{s_3}{s}f(ir,iz) + \frac{s_4}{s}f(ir,iz+1) + \frac{s_1}{s}f(ir+1,iz+1) + \frac{s_2}{s}f(ir+1,iz), \tag{9}$$

where $s_1, s_2, s_3, s_4$ are the area of the rectangle enclosed by the particle and the neighboring grid points respectively shown in Figure 1. The area of a grid is $s = s_1 + s_2 + s_3 + s_4 = dr \times dz$. When finding the particle density $n_i$, integrating the particle sources to grids is actually the inverse process of interpolation with same coefficients. To suppress numerical dispersion and filter short waves, we smooth the density by averaging over neighboring grids

$$n_i(ir_0, iz_0) = \frac{1}{(2N_{sm}+1)(4N_{sm}+1)} \sum_{ir=ir_0-N_{sm}}^{ir_0+N_{sm}} \sum_{iz=iz_0-2N_{sm}}^{iz_0+2N_{sm}} n_i(ir, iz). \quad (10)$$

Smoothing are performed twice in the HYPIC program, with the smoothing parameter $N_{sm}$ taken as 2 and 3, respectively. For particle injection, we generate an initial position distribution of ions satisfying a Gaussian distribution and an initial velocity distribution satisfying a Maxwellian distribution by random numbers. When $N_s$ particles run out of the boundary, the information of these $N_s$ particles is thrown away. At the next time step, new $N_s$ particles will be injected, satisfying a Gaussian distribution for position and a Maxwellian distribution for velocity.

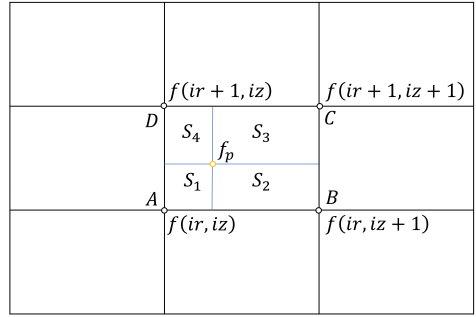

Figure 1. The PIC bilinear interpolation interpretation, i.e., area $s_1, s_2, s_3, s_4$ to grid point C, D, A, B respectively.

We use the MCC method proposed by Nanbu [18], based on the idea of grouping small-angle binary collisions in plasmas into a unique binary collision with a large scattering angle. Thus, the time step can be increased and the computational time can be greatly reduced. Since it is described carefully enough in reference [18], we will not repeat it.

We use the Yee grid [23] to discretize the fluctuation equations (5-6). Quantities not on the grids are averaged over neighboring grids. The final discretized equations can be written in the form of a system of linear equations and solved by matrix inversion. The HYPIC program solves for 3D $(r, \theta, z)$ EM field distributions excited by single-loop antennas [24], helical antennas [25], half-turn antennas [5, 9], and Nagoya antennas [26] in non-uniform magnetic fields and non-uniform plasma. Usually the plasma density and electron temperature are slowly varying with time, and thus we can solve the equations (5-6) once every 10 RF cycles to save computational resources. In fact, the EM module in HYPIC and in PHD v2.0 [27] is the same.

Introducing the magnetic vector potential $\boldsymbol{A}$,

$$\boldsymbol{B}_0 = \nabla \times \boldsymbol{A}, \tag{11}$$

the equation (2) becomes

$$\nabla(\nabla \cdot \boldsymbol{A}) - \nabla^2 \boldsymbol{A} = \mu \boldsymbol{J}_c. \tag{12}$$

Considering the magnetic field generated by the coils in the cylindrical coordinate system, i.e., $\boldsymbol{J}_c = J_\theta \boldsymbol{e}_\theta$, we have

$$-\left(\frac{\partial^2 A_\theta}{\partial r^2} + \frac{\partial^2 A_\theta}{\partial z^2} + \frac{1}{r}\frac{\partial A_\theta}{\partial r} - \frac{A_\theta}{r^2}\right) = \mu J_\theta. \tag{13}$$

The equation (13) are discretized by central differences and solved by Gaussian iteration. The iteration format without ferromagnetic medium is

$$\left(\frac{2}{\Delta r^2} + \frac{2}{\Delta z^2} + \frac{1}{r^2}\right) A_\theta(ir, iz) = \mu_0 J_\theta +$$
$$\left(\frac{A_\theta(ir+1, iz) + A_\theta(ir-1, iz)}{\Delta r^2} + \frac{A_\theta(ir, iz+1) + A_\theta(ir, iz-1)}{\Delta z^2} + \frac{A_\theta(ir+1, iz) - A_\theta(ir-1, iz)}{2r\Delta r}\right). \tag{14}$$

For the case with a ferromagnetic medium, the iteration format can be found in reference [19]. Finally, we can obtain the magnetic field by

$$\boldsymbol{B} = -\frac{\partial A_\theta}{\partial z} \boldsymbol{e}_r + \left(\frac{\partial A_\theta}{\partial r} + \frac{A_\theta}{r}\right) \boldsymbol{e}_z. \tag{15}$$

The ES field, magnetic field and density are calculated and interpolated in 2D cylindrical coordinate system $(r, z)$. The motion of the particles is calculated in 3D Cartesian coordinate system $(x, y, z)$ and the induced electric field is calculated in 3D cylindrical coordinate system $(r, \theta, z)$. Taken together, HYPIC is a 2D $(r, z)$ EM PIC-MCC program able to fast simulate the antenna-plasma interactions and ICRE processes in the cylindrical coordinate system. In addition, the HYPIC program has a good modularity for the background magnetic field, the motion of the particles, the RF antenna-plasma interaction, and the MCC, so that the user can easily use one or more of these functions. Unless otherwise specified, the parameters in this paper are shown in Table 1.

Table 1. Default parameter settings.

| Parameters | Value |
|---|---|
| atomic mass number | 1 (hydrogen ion) |
| Number of ions $N_p$ | $10^5$ |
| Radio frequency $f$ | 13.56 MHz |
| Radial grids $N_r$ | 101 |
| Axial grids $N_z$ | 101 |
| Initial electron temperature $T_{e0}$ | 3 eV |
| Initial ion temperature $T_{i0}$ | 3 eV |
| Radial boundary of the plasma $r_p$ | 0.07 m |
| macroparticle number $N_m$ | $3.2 \times 10^{11}$ |

## 2.3 Benchmarks

For the verification of the background magnetic field, we compare the numerical results with the analytical solution. The problem to be solved is the axial distribution of the magnetic field generated by an ideal coil, with analytical solution

$$B_0(z) = \frac{\mu_0 I_0 R_c^2}{2(R_c^2 + (z - Z_c)^2)^{\frac{3}{2}}}, \qquad (16)$$

where the radius and axial position of the coil are $R_c = 0.1$ m, $Z_c = 0.5$ m, and the current of the coil is $I_0 = 10^5$ A. In HYPIC calculation, we set a coil with radial width 0.01 m and the axial width 0.02 m. The computational region is $r \in [0, 0.3]$ m, $z \in [-0.5, 1.5]$ m and the number of grids is $N_r \times N_z = 301 \times 2001$. Figure 2 shows the results calculated by the HYPIC program and the analytical solution for the ideal coil, and it can be seen that they are in general agreement, with a maximum difference of 2.4%. This small difference may be due to the non-ideal coil in HYPIC calculation. Therefore, the background magnetic field module in the HYPIC program is reliable.

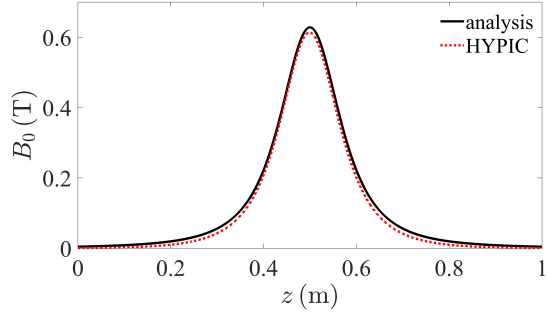

Figure 2. The axial magnetic field generated by coils, where the black solid line is the analytical solution of the ideal coil and the red dashed line is the result of the HYPIC program.

The solution of the motion equation (1) and the conservation of energy are verified next. The evolution of the mean kinetic energy in the system with time is discussed here by considering only the Lorentz force (magnetic field) and ignoring the electric field force and collisions. Since the Lorentz force does not do work, the kinetic energy of the particles should be conserved. To avoid the loss of ions, we adopt a periodic boundary condition in the axial boundaries and a reflection boundary condition in the radial boundary. The magnetic field of magnetic mirror used here is shown in Figure 3. The number of particles is $N_p = 10^4$. We set the initial velocities $v_{x0} = 10v_{y0} = 10v_{z0}$ in order to make the particles as constrained as possible by the magnetic mirror. Figure 4 shows the evolution of the mean kinetic energy of all particles $E_k$ over time, and it can be seen that the smaller the time step, the better the energy conservation. In particular, the energy loss rate $(1 - E_k/E_{k0})$ at $t = 0.01$ s is $1.07\%$ in the case of $dt = 10^{-2} T_{rf}$. Our calculations are usually within 0.01 s, so we default to $dt = 10^{-2} T_{rf}$. In addition, although the RK4 method is in explicit form, it is dissipative in nature.

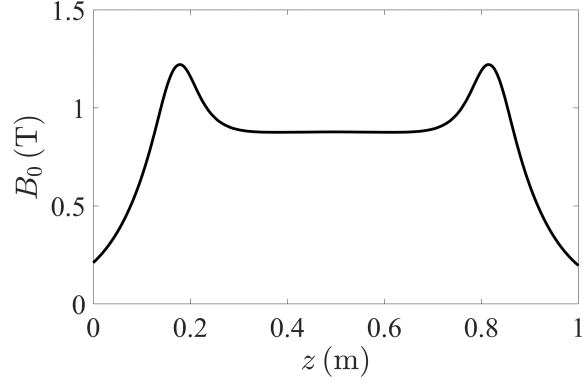

Figure 3. Distribution of magnetic field with axial position at $r = 0.02$ m.

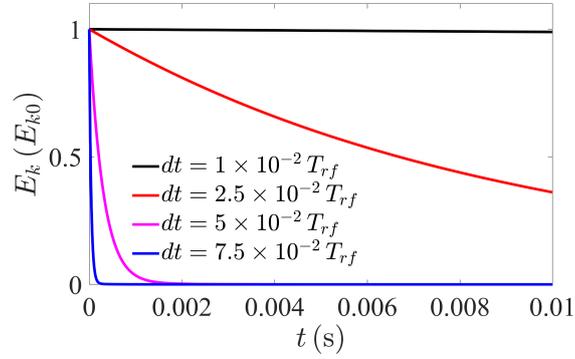

Figure 4. The evolution of the mean kinetic energy of all particles $E_k$ over time, where only the Lorentz force is considered and $E_{k0}$ is the initial mean kinetic energy.

The MCC module is then verified. Two questions are discussed here, including ion velocity reaching isotropy from anisotropy through i-i collisions, and ions and electrons reaching thermal equilibrium through collisions. We set the initial temperature $T_x = 1.3T_y = 1.3T_z$ in the first question (i-i collisions). The total ion temperature is $\left(= \frac{1}{3}T_x + \frac{2}{3}T_y\right)$, and the analytical solution $\Delta T (= T_x - T_y)$ [18, 28] is

$$\Delta T_{xy} = (\Delta T_{xy})_0 \exp\left(-\frac{8}{5\sqrt{2}}\frac{t}{\tau_0}\right), \qquad (17)$$

where the collision time (reference time) $\tau_0 = 8\pi\sqrt{2m_i}\varepsilon_0^2 T_i^{3/2}/(n_i e^4 \ln\Lambda)$, $T_i$ in eV, the density of the ions $n_i$, and the initial temperature difference $(\Delta T_{xy})_0$. The collision time step is taken as $dt_{mcc} = 10^{-2}\tau_0$. Figure 5 (a) shows that the simulation results are in good agreement with the analytical solution.

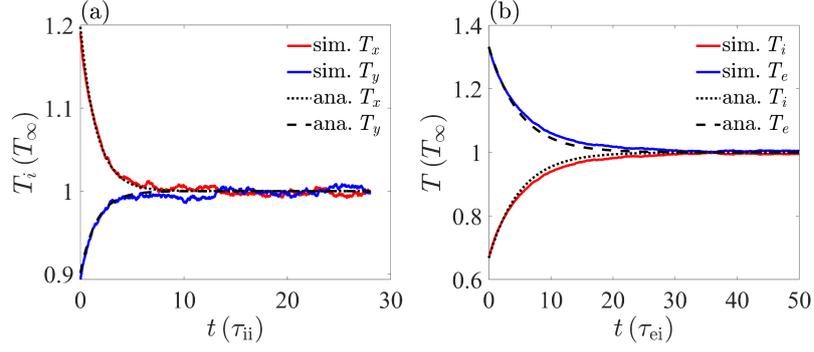

Figure 5. The (a) i-i collision and (b) e-i collision relaxation process.

Then, we take the initial temperature $T_{e0} = 2T_{i0}$ in the second question (e-i collision relaxation process) with e-i, i-e, e-e, i-i collisions considered. In order to accelerate the relaxation process, the electron mass is taken to be $m_e = m_i/4$. According to the reference [28], the temperature difference $\Delta T_{ei}$ ($T_e - T_i$) is solved analytically as

$$\Delta T_{ei} = (\Delta T_{ei})_0 \exp(-2\nu_{eq}t), \tag{18}$$

with $\nu_{eq} = 8\nu_0 m_e (1 + m_e T_i/(m_i T_e))^{-\frac{3}{2}}/(3m_i\sqrt{\pi})$, $\nu_0 = n_i e^4 \ln \Lambda / \left(8\pi\sqrt{2m_e}\varepsilon_0^2 T_e^{3/2}\right)$. The collision time step is taken as $dt_{mcc} = 10^{-2}/\nu_0$. Figure 5 (b) shows the comparison of the simulation with the analysis, indicating the convergence times of the two results being close. In fact, these two questions and the simulation results in Figure 5 are similar to those in reference [19]. Overall, the HYPIC simulation are in good agreement with the analysis, indicating the MCC module being verified.

The HYPIC transport module is verified by the magnetohydrodynamic (MHD) program PHD v1.1 [21]. The problem discussed is to solve for the steady-state plasma density given the plasma source $S = S_0 \exp\left(-\frac{r^2}{2r_w^2}\right)\exp\left(-\frac{(z-z_c)^2}{2z_w^2}\right)$, where radial width, axial width and axial center position are $r_w = 0.025 \text{ m}, z_w = 0.1 \text{ m}, z_c = 0.5 \text{ m}$ respectively. The effects of ES fields, magnetic fields and collisions are considered. The transport model for the PHD program [21] is

$$\frac{\partial n}{\partial t} + \nabla \cdot \boldsymbol{\Gamma} = S, \tag{19}$$

$$n = n_i = n_e, \tag{20}$$

$$\boldsymbol{\Gamma} = -D_\perp \nabla_\perp n \boldsymbol{e}_\perp - D_a \nabla_\parallel n \boldsymbol{e}_\parallel. \tag{21}$$

The electron temperature is fixed at $T_e = 3$ eV.

The case of $B_0 = 0$ with $S_0 = 1.68 \times 10^{25}$ m$^{-3}$s$^{-1}$ is discussed first. Figure 6 shows the steady-state density $n_i$ obtained by the HYPIC and PHD simulations. The plasma density distributions and values are broadly similar for both. The density of HYPIC is slightly wider than that of PHD, probably due to the smoothness in HYPIC.

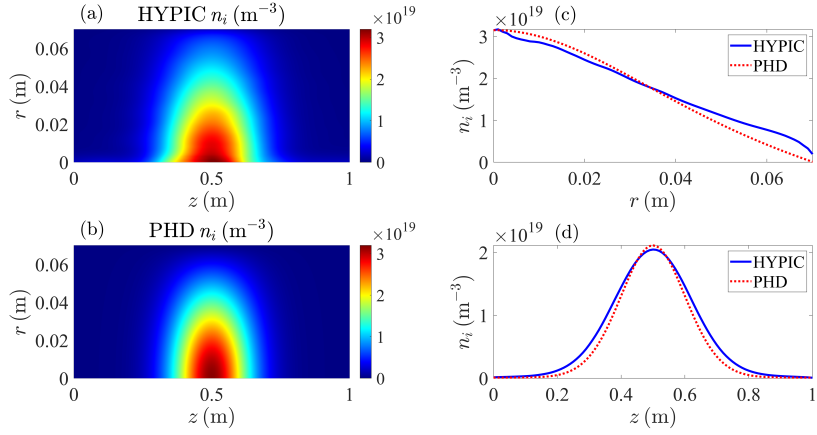

Figure 6. Steady-state ion density distributions simulated by (a) the HYPIC program and (b) the PHD program with $B_0 = 0$. (c) The radial ion density at $z = 0.5$ m and (d) the axial the ion density at $r = 0.03$ m.

The case of $B_0 = 0.4$ T with $S_0 = 1.61 \times 10^{23}$ m$^{-3}$s$^{-1}$ is then discussed. The plasma density distributions and values are also similar for HYPIC and PHD, shown in Figure 7. The difference between the two may also be caused by the smoothness in HYPIC.

Overall, HYPIC and PHD give similar results, verifying HYPIC's transport module.

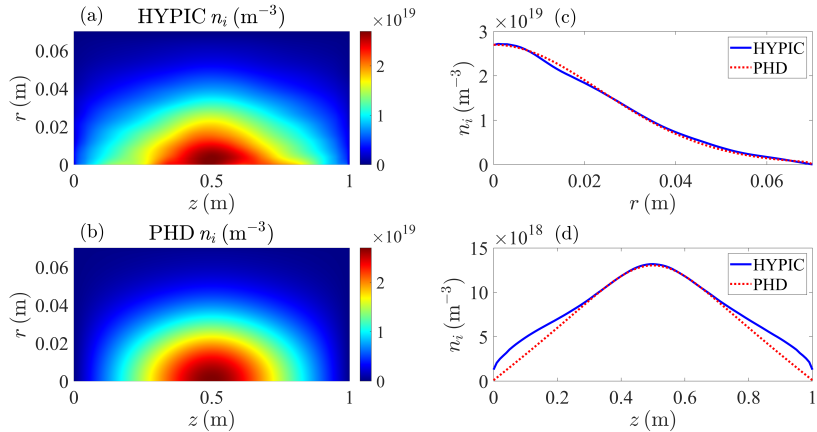

Figure 7. Steady-state ion density distributions simulated by (a) the HYPIC program and (b) the PHD program with $B_0 = 0.4$ T. (c) The radial ion density at $z = 0.5$ m and (d) the axial the ion density

at $r = 0.03$ m.

Finally, the EM modules in HYPIC and PHD v1.1 [21] are compared. Unlike the FDFD in HYPIC, PHD uses FDTD method. The induced electric field $\boldsymbol{E}_{mf}$ generated by a single loop antenna is calculated. The position of the antenna is $(r_a, z_a) = (0.08, 0.5)$ m with the radial width 0.01 m and axial width 0.04 m. The plasma density is Gaussian distributed with $n_e = n_{e0} \exp\left(-\frac{r^2}{2r_w^2}\right) \exp\left(-\frac{(z-z_c)^2}{2z_w^2}\right)$, where $r_w = 0.02$ m, $z_w = 0.2$ m, $z_c = 0.5$ m, $n_{e0} = 10^{19}$ m$^{-3}$. The radius of the plasma is $r_p = 0.07$ m. The ideal conductor boundary conditions are used for both axial and radial boundaries. The uniform background magnetic field is taken, i.e., $\boldsymbol{B}_0 = B_0 \boldsymbol{e}_z, B_0 = 0.4$ T. The amplitude of $\boldsymbol{E}_{mf}$ obtained by HYPIC and PHD is shown in Figure 8. It can be seen that the radial electric fields $E_{mfr}$ are both centrally peaked, but $E_{mfr}$ in PHD is narrower than that in HYPIC, probably due to HYPIC optimization $(\nabla \cdot \boldsymbol{E}_{mf} = 0)$ in the vacuum region. The azimuthal and axial electric field distributions of HYPIC and PHD are very close to each other both in terms of values and shapes. Overall, HYPICs EM module is reliable.

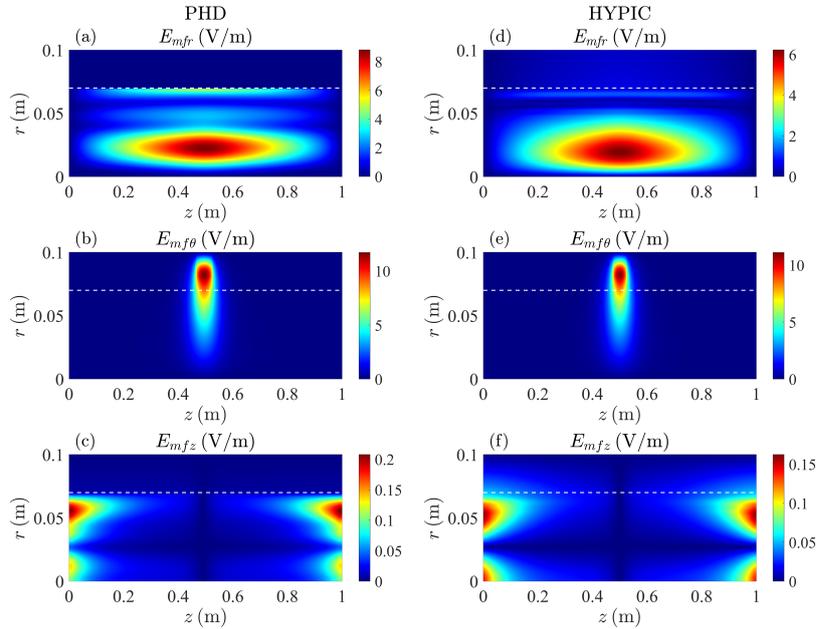

Figure 8. The (a) radial electric field $E_{mfr}$, (b) azimuthal electric field $E_{mf\theta}$, (c) axial electric field $E_{mfz}$ calculated by HYPIC and the (d) radial electric field $E_{mfr}$, (e) azimuthal electric field $E_{mf\theta}$, (f) axial electric field $E_{mfz}$ calculated by PHD.

To make a small conclusion, we have carried out detail verifications of HYPIC and confirmed that the HYPIC program is reliable.

## 3 Application

Setting different conditions such as background magnetic field, antenna, plasma density, configuration, etc., the HYPIC program can be used to fast simulate the antenna-plasma interactions and ICRE processes in a variety of linear devices, such as magnetic mirrors, high-power electric propulsion, and FRC. The ICRE in magnetic mirrors as an example is shown next.

### 3.1 Simulation configuration

We have designed a magnetic mirror shown in Figure 9, consisting of three coils, the main coil, the left coil, and the right coil. The main coil generates a relatively uniform background magnetic field, and the left and right coils generate the stronger confining magnetic field, shown in Figure 9 (a). The position coordinates $(r_1, r_2, z_1, z_2)$ of the main, left and right coils are $(0.1, 0.12, 0.2, 0.8)$ m, $(0.07, 0.1, 0.15, 0.2)$ m, $(0.07, 0.1, 0.8, 0.85)$ m, and the numbers of ampere-turns are $4.50 \times 10^5$ A, $1.31 \times 10^5$ A, $1.31 \times 10^5$ A, respectively. The magnetic lines are shown in Figure 9 (a) and the magnitude of the magnetic field is shown in Figure 9 (b). The coordinates $(r_1, r_2, z_1, z_2)$ of the single loop antenna are $(0.08, 0.09, 0.48, 0.52)$ m. The uniform background magnetic field near the antenna is intended to ensure high energy coupling efficiency [9]. The RF frequency is $f = 13.56$ MHz and the current amplitude is $I_{rf} = 10$ A.

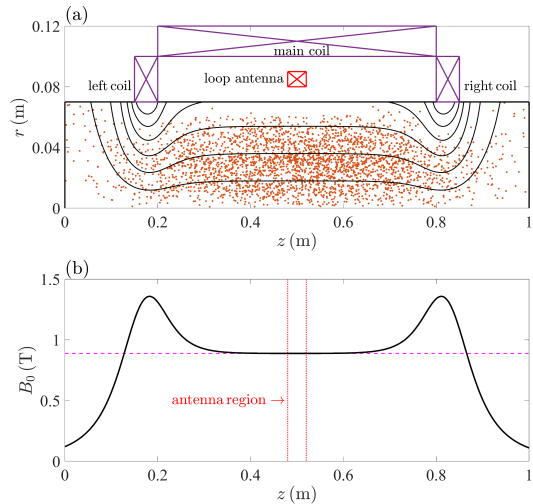

Figure 9. (a) Schematic of the simulated configuration, where the coils generating the background

magnetic field are marked in purple. The thick black solid lines are the plasma boundary, and only the magnetic lines (thin black solid lines) in the plasma region are shown. The ions are represented by the brown dots. (b) Variation of the magnetic field at $r = 0.02$ m, with the horizontal dashed line representing the resonant magnetic field.

### 3.2 Simulation results

Figure 10 shows the evolution of the mean kinetic energy of ions $E_k$ and electron temperature $T_e$ with time, indicating that the ion kinetic energy increases firstly and then decreases and gradually stabilizes. Another important information of $E_{kx} \approx E_{ky} \gg E_{kz}$ indicates that the ICRE mainly increases the cyclotron kinetic energy $(E_{kx} + E_{ky})$. It is difficult for ions to reach thermal equilibrium $(E_{kx} \approx E_{ky} \approx E_{kz})$ through collisions, and that is why we usually call it ICRE (without sufficient collision) instead of ICRH (with sufficient collision). The electron temperature is significantly less than the ion kinetic energy, indicating that electrons and ions also have difficulty in reaching thermal equilibrium through collisions.

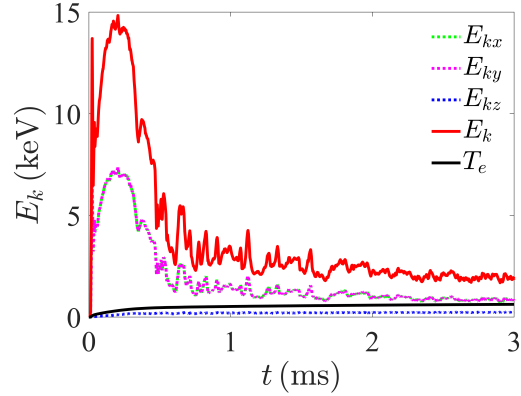

Figure 10. Evolution of the mean kinetic energy of ions and electron temperature with time.

Figure 11 shows the distribution of $E_k$ along the axial position at $t = 3$ ms. $E_{kx} + E_{ky} \gg E_{kz}$ near the antenna $(0.3 \text{ m} < z < 0.7 \text{ m})$ also indicates that the ICRE mainly increases the cyclotron kinetic energy. At $z < 0.2$ m and $z > 0.8$ m, the cyclotron kinetic energy decreases while the parallel kinetic energy increases rapidly due to two reasons of (1) the axial ES field $E_{sz}$ accelerating the ions, and (2) the evanescent magnetic field converting the cyclotron kinetic energy $(E_{kx} + E_{ky})$ to the parallel kinetic energy $E_{kz}$ due to the conservation of magnetic moments.

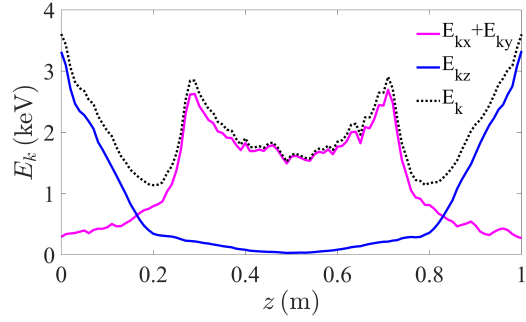

Figure 11. Distribution of the mean kinetic energy of ions along the axial position at $t = 3$ ms.

Figure 12 shows the kinetic energy distribution of the ions along the axial direction at $t = 3$ ms. The $E_{kx}$ and $E_{ky}$ have a good symmetry about $E_k = 0$. The kinetic energy is widely distributed in the region of the magnetic mirror confinement $(0.3 < z < 0.7 \text{ m})$, with many energetic ions. In the region of $z < 0.2$ m and $z > 0.8$ m, most of the ions have axial velocities pointing toward the boundary.

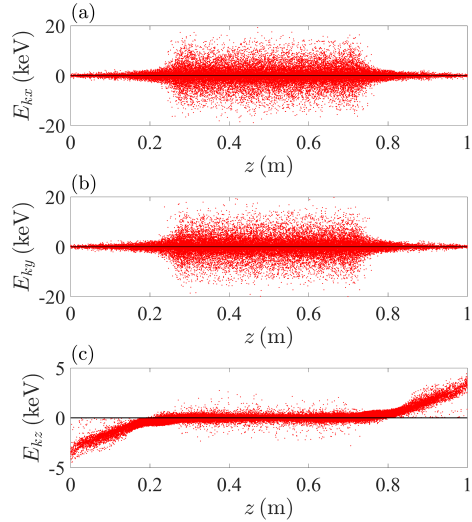

Figure 12. The ion kinetic energy distribution (a)$E_{kx}$, (b)$E_{ky}$ and (c)$E_{kz}$ along the axial direction at $t = 3$ ms, with positive and negative representing their velocity directions.

Figure 13 shows the 2D $(r, z)$ distribution of ion density $n_i$, ion kinetic energy $E_k$, radial ES field $E_{sr}$, axial ES field $E_{sz}$, radial induced electric field $E_{mfr}$, and azimuthal induced electric field $E_{mf\theta}$ $t = 3$ ms. The density is concentrated in the region of the magnetic mirror confinement, where the ions are accelerated up to about 3 keV. The radial ES field has large values mainly at the radial boundary, where the plasma density gradient is large. Both radial and axial ES fields have large values near the plasma boundary. After careful analysis, it is found that the phase difference of time evolution between the radial ion velocities $v_r$ and the radial induced electric field $E_{mfr}$ is close to 0. Thus, it can be obtained that the $E_{mfr}$ dominates the ICRE here. Unlike the azimuthal induced electric field $E_{mf\theta}$ accelerating ions near the radial boundary in low plasma density [9], the $E_{mfr}$ can accelerate the core ions in higher plasma density, beneficial to confinement. In fact, the effective ICRE is expected to solve the end-loss problem of the magnetic mirror fusion. The methods are that ICRE can rapidly increase the cyclotron kinetic energy when the ions entered the loss cone by collisions, leading the ions out of the loss cone and into the confinement state. The study of confinement and energization in magnetic mirror is not the subject of this work and will be discussed in detail elsewhere.

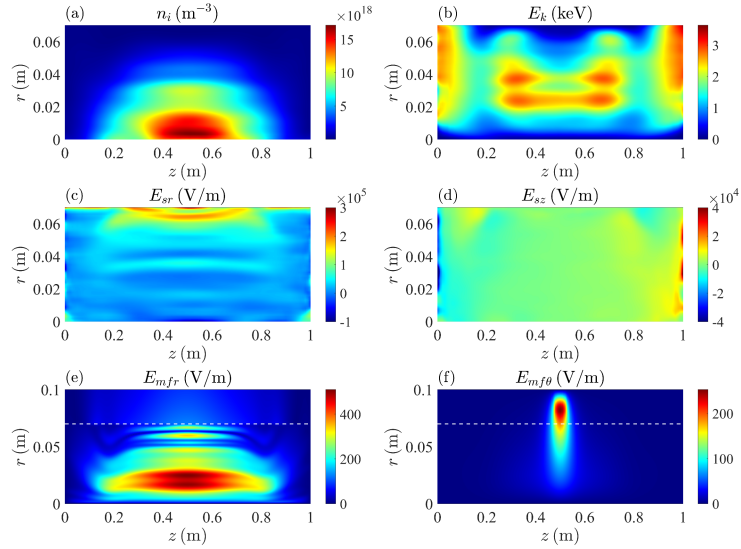

Figure 13. Two-dimensional distribution of (a) ion density $n_i$, (b) ion kinetic energy $E_k$, (c) radial ES field $E_{sr}$, (d) axial ES field $E_{sz}$, (e) radial induced electric field $E_{mfr}$, and (f) azimuthal induced electric field $E_{mf\theta}$ $t = 3$ ms.

## 4 Conclusion and discussion

To summarize, we have developed a 2D $(r, z)$ EM PIC-MCC program in the cylindrical coordinate system, named HYPIC, able to fast simulate the antenna-plasma interactions and the ICRE processes in linear devices such as high-power electric propulsion, magnetic mirror, and FRC, etc. We provide a detailed verification of each module and an example of the ICRE in magnetic mirror. The main advantages of the HYPIC program are inclusion of (1) the kinetic effects of ions, (2) the ES and EM effects, (3) the collisions between ions and electrons, and (4) with a small computation, e.g., the example given in section 3 consuming 20 hours by one thread. Although the collisional effects on the electron temperature is considered, the spatial distribution of the electron temperature and the heating effect of RF waves are not considered. The electron temperature evolution equation will be solved in the next work.

## Author contributions

**Mingyang Wu**: Conceptualization, Data curation, Formal analysis, Investigation, Methodology, Software, Validation, Writing - original draft, and Writing - review & editing. **Andong Xu**: Data curation, Formal analysis, Investigation, Methodology, Software. **Chijie Xiao**: Supervision, Funding acquisition, Project administration.

## Declaration of competing interest

The authors declare that they have no known competing financial interests or personal relationships that could have appeared to influence the work reported in this paper.

## Data availability

Data will be made available on request.

## Acknowledgments

We would like to acknowledge Ting Gao for useful discussions. This work was supported by the National Key Research and Development Program of China (No. 2022YFA1604600).